# The conceptual design of 100-kA pulsed magnetic field generator for magnetized high-energy-density plasma experiments.

R.V. Shapovalov[1,2], R. B. Spielman[2], G. Brent[2], and P.-A. Gourdain[1,2]

[1]Department of Physics and Astronomy, University of Rochester, Rochester, NY 14627, USA
[2]Laboratory for Laser Energetics, University of Rochester, Rochester, NY 14623, USA

**Abstract.** This paper presents the conceptual design of a high-voltage pulser intended to generate 30-T magnetic fields for magneto-inertial fusion experiments at the OMEGA facility. The pulser uses a custom capacitor bank and two externally triggered spark gaps to drive a multi-turn coil. This new high-voltage pulser is capable of storing 10 times more energy than the previous system, using a higher charge voltage (from 20 to 30 kV) and a larger capacitance (from 1 µF to 5 µF). Circuit simulations shows that this pulser can deliver 100 kA into a 60-nH, 14-mΩ coil with a rise time of 1 µs. For a coil with 2 turns with an average coil diameter of 7.8 mm, this current translates into a 32-T peak magnetic field at coil center. This is a factor of three increase in the peak magnetic field compared to the present generator magnetic field capabilities.

Keywords: **Pulsed power driver; High magnetic field; High energy density physics; Conceptual design.**

Introduction

Magnetized plasmas and magnetized, inertial-confinement-fusion (ICF) experiments are an important milestone in fusion energy research. Pioneered on the Omega laser[1], this research integrates basic plasma physics, transport in magnetized plasmas, alpha-particle energy deposition in ICF capsules, magneto-inertial-fusion concepts[2,3,4,5,6]. It also enables scientists to tackle more fundamental physics questions by studying positron focusing[7], astrophysical shocks[8,9,10], magnetized plasma jets, and magnetic reconnection[11]. However, the engineering constraints (volume and weight) in getting a high-current pulser into the Omega target chamber fundamentally limit the energy that is available for the production of magnetic fields.

MIFEDS-2 is a second-generation high-voltage pulser/coil system currently used for magnetized high-energy plasma experiments at the OMEGA and OMEGA-EP laser facilities[12]. The present system combines two, 0.5-µF (General Atomics) capacitors charged to 20 kV with a single, vacuum spark gap (PerkinElmer GP-14B) located on the high-voltage (HV) side of the unit. At this charging voltage, the MIFEDS-2 capacitors can store up to 200 J of energy. About 20% of the energy stored in the capacitors can be used to generate a magnetic field when using a typical 60-nH coil as the load. The pulser internal inductance is measured to be 139 nH and the pulser internal resistance is about 130 mΩ. Even for a low-inductance coil, these driver values limit the total current to about 40 kA, making it difficult to reach magnetic field larger than 20 T.

In this paper, we present the design of MIFEDS-3, a new high-voltage pulser capable of generating magnetic fields greater than 30 T. The general view of the MIFEDS-3 pulser is shown in Figure 1. The pulser is comprised of two, custom 2.5-µF capacitors (General Atomics or Scientific Applications & Research Associates, Inc.) connected in parallel and charged to a maximum voltage of 30 kV. At this charging voltage, the stored capacitor energy is 2.25 kJ, which is 10 times larger than the energy available in the MIFEDS-2 system. This gain in energy is achieved by increasing the physical size of the capacitors, requiring the HV power supply to be moved outside of the Omega ten-inch-manipulator (TIM). The two pulser capacitors are each discharged through a single spark gap (PerkinElmer GP-14B). These two spark gaps are effectively connected in parallel on the ground side of TIM. The estimated internal pulser inductance is 90 nH, which is 1.5X smaller than the present MIFEDS-2 pulser design. The internal resistance is estimated to be less than or equal to 140 mΩ.

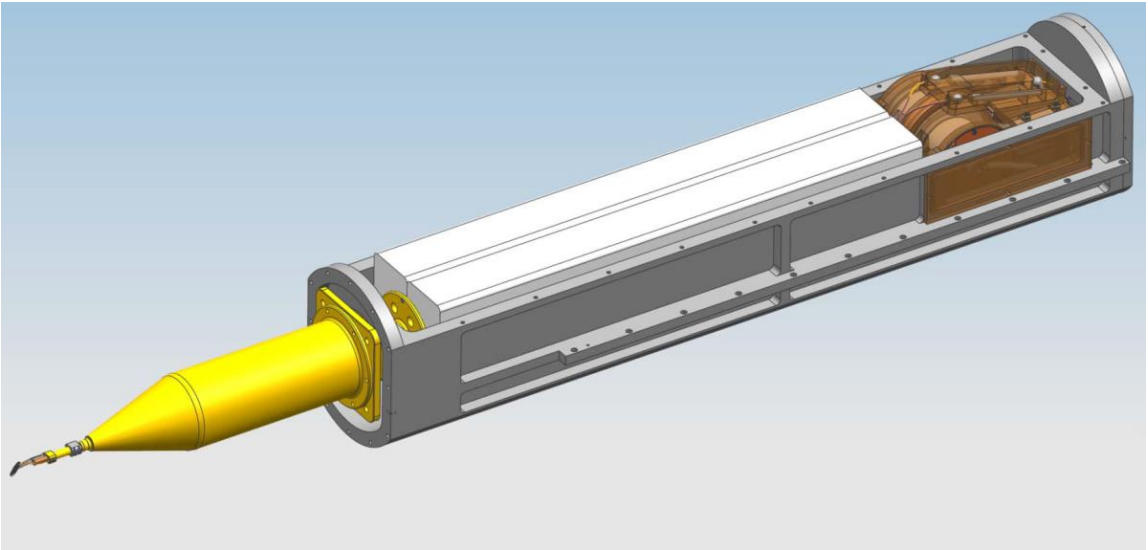

Figure 1 MIFEDS-III design view.

We first present the detailed MIFEDS-3 pulser circuit simulations into some average 60-nH, 14-mΩ coil. We next discuss how the driver parameters scale with the coil inductance. And finally, we present the coil magnetic field calculations for 1-, 2-, 3-, and 4-turn coils and compare them with the MIFEDS-2 pulser. The driver is capable of suppling a maximum current of 100 kA. We expect that the MIFEDS-3 pulser will generate more than 30-T peak magnetic field inside simple, 2-turn magnetic coil.

**Pulser Circuit Modeling.**

We first present detailed circuit simulations into a typical MIFEDS coil. For consistency, we assume a 2-turn coil with average coil diameter of 7.8 mm that is made of 24-AWG (American wire gauge) copper-insulated wire. The 24-AWG wire has an outer diameter of 0.762 mm and the copper core diameter of 0.508 mm. For a single turn, the coil inductance can be approximated by[13] $L \approx \mu_0 R[ln(8R/r) - 1.75]$, where $\mu_0$ is permeability of vacuum, R is the coil radius, and r is the wire radius. The total wire length equals to $l = N\pi d + 12\ cm$, where N is the number of turns, d is the coil diameter, and 12 cm is an extra length of the coil wire legs. For $N = 2$ turn coil, the coil inductance ($N^2$ times larger) is 60 nH and the coil resistance is 14 mΩ.

In following simulations, we follow the matched load approach, $R = \sqrt{L/C}$. This is an ideal case for the high-current driver when one wants simultaneously to maximize the load current and minimize the voltage reversal (i.e. ringing) inside the system[14]. The total pulser inductance (internal value plus coil value) is equal to 150 nH, and the total pulser resistance is matched to $\sqrt{L/C}$ and equals to 170 mΩ. The matched value is slightly larger than the total pulser resistance (internal value plus coil value) of 154 mΩ and can be compensated by installing in-series high-resistance wire, if desired. It is important to note that the MIFEDS-3 internal resistance, 140 mΩ, is, by design, close to the matched $\sqrt{L/C}$ pulser value, 170 mΩ, maximizing the load peak current and preventing the unwanted late-time current oscillations. We assume, for simplicity, no change in the wire resistance during the current pulse. This should be a good assumption, as the coil resistance is a small portion of the total pulser resistance. All simulations are performed with the SCREAMER[15], a special-purpose pulsed-power circuit simulations tool.

The resulting current and voltage waveforms are presented in Figure *2*. A peak current of 95 kA is reached 1.05 µs after the start of the current waveform (the quarter-cycle time). The maximum coil voltage is 12 kV at the beginning of the coil pulse, and the minimum voltage is -2.8 kV at about 2.4 µs after the pulse starts. The resistive part of the voltage drop, thin dashed line, reaches its maximum value of 1.9 kV at 1.05 µs, and the inductive part of the voltage, thin dot-dashed line, follows closely the total voltage applied to the coil. We observe that both the positive and the negative peak currents are well inside the capacitor's safety margins. The negative reversal current is -16 kA at a time of ~ 4 µs with minimal current oscillations afterworld. The presented pulser design, being very close to the matched load circuit, optimizes the coil peak current but limits the pulser current oscillations at later times.

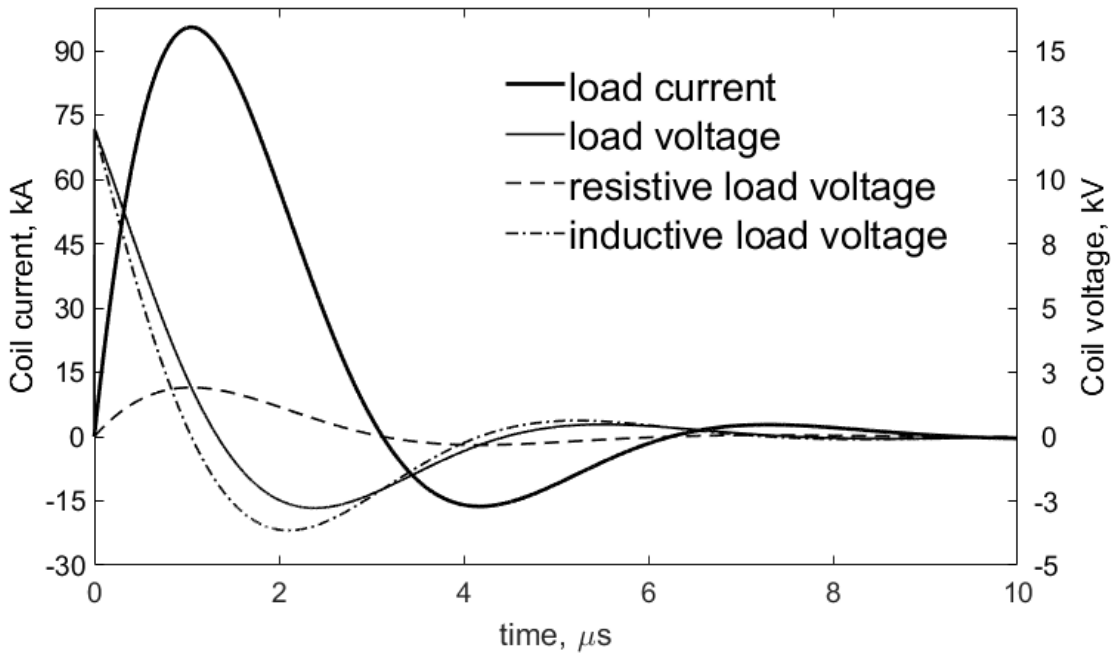

Figure 2. Current and voltages waveform into a matched 60-nH coil are shown. The thick solid line is the total load current, thin solid line is load total load voltage, the dashed line is the load resistive voltage, and the dot-dashed line is the load inductive voltage.

It is interesting to compare the stored electrical energy of the pulser with the electrical energy delivered to the coil at the time of peak current, 1.05 µs. The total energy stored in the pulser is 2.25 kJ. The total energy delivered to the 60-nH coil, including its resistive and inductive parts, is plotted in Figure *3*. The peak energy delivered to the coil is ~400 J or ~ 18% of the initial energy stored in the capacitor. The resistive energy heats the coil and by the time of the current peak only 42% (0.11 kJ/0.26 kJ) of the total resistive contribution is deposited in the coil. The inductive part of the energy, as expected, reaches its maximum value of 0.27 kJ at the time of the peak current. The MIFEDS-3 pulser delivers 12% of the initially stored electrical energy into the magnetic energy of the coil. It is tempting to increase the time-to-peak of the current (larger capacitors) to increase the total energy stored inside the magnetic field, but longer current rise times will result in increased motion of the coil wires, preventing the coil from reaching its maximum value of peak magnetic field. Keeping this current rise below 1-2 µs gives an optimal coupling between the initial stored energy and the magnetic energy. Current rise times shorter than 1-µs result in excessive coil voltages when the peak current is increased beyond MIFEDS-2 currents.

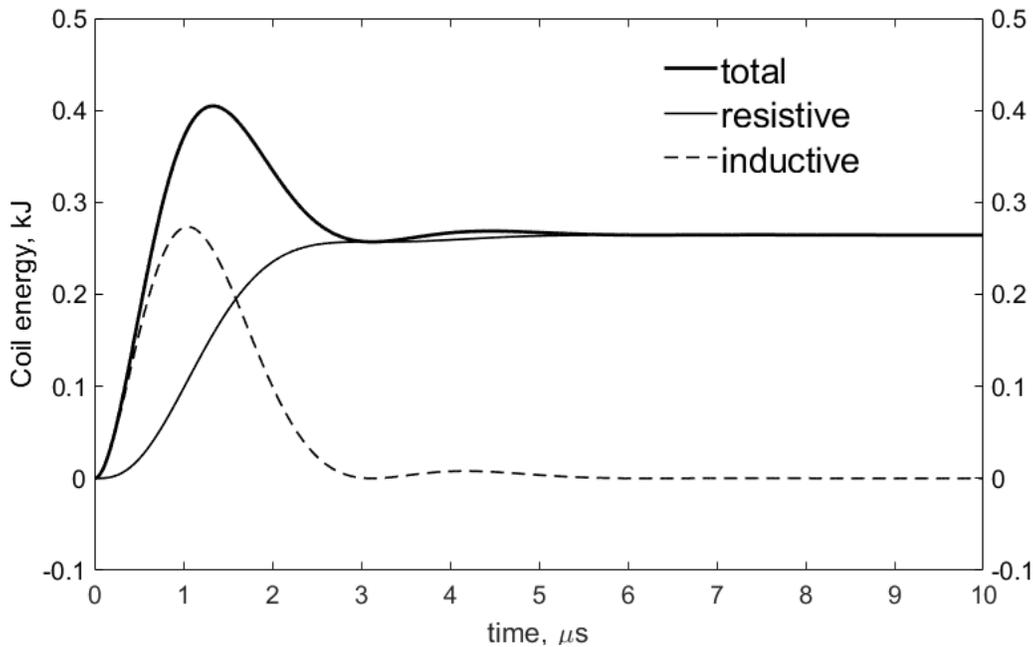

Figure 3 The energy deposition into a matched 60-nH coil. The thick solid line is the total energy delivered to the coil, the thin solid line is its resistive part, and the thin dotted line is its inductive part.

Other experiments may require different coils and we are interested in the dependence of the pulser output parameters on coil inductance. The most important pulser parameters affecting the resulting coil magnetic field are the pulser peak current and the time-to-peak current. A scan of these pulser parameters as a function of coil inductance is presented in Figure *4*. Each data point is a separate SCREAMER run with the coil inductance value from the horizontal axis. The total driver inductance is, as before, the sum of the pulser internal inductance plus the inductance of the corresponding coil. However, we do not assume the matched, $\sqrt{L/C}$, resistance value as we did earlier for 2-turn, 60-nH coil. Instead, we assume the constant coil resistance of 20 mΩ, and the constant total pulser resistance of 160 mΩ (140 mΩ + 20 mΩ) for all SCREAMER runs for all pulser coils. A more accurate estimate of the coil resistance will be possible when the coil wire length is better known.

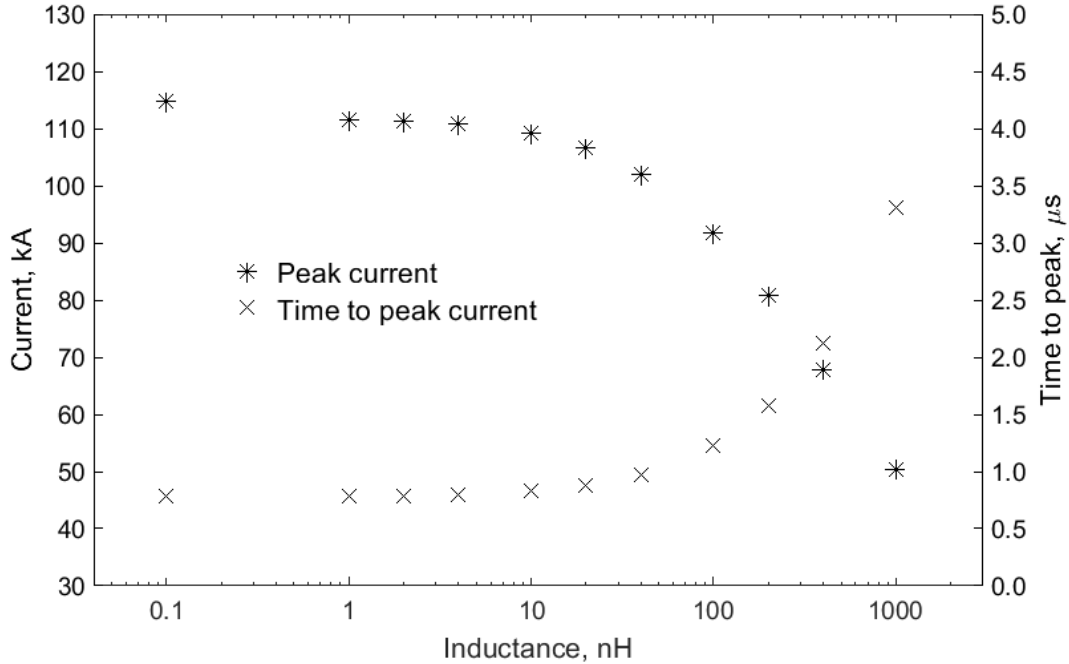

Figure 4. MIFEDS-3 coil peak current and time-to-peak current as a function of the coil inductances. The coil resistance is assumed to be 20 mΩ for all different runs.

As expected, the pulser peak currents and time-to-peak currents are monotonic functions of the coil inductance. As we increase the coil inductance from 0.1 nH to 100 nH, the peak currents decrease from 115 kA down to 50 kA, and time-to-peak currents increase from 0.8 µs up to 3.3 µs. For low-inductance coils below 40 nH, the peak currents are slightly above the operational pulser limit of 100 kA (switch and capacitor limited). For the high-inductance coils above 400 nH, the time-to-peak currents become larger than 2 µs, and the coil performance will degrade because the increased wire motion. These calculations suggest that the optimal pulser performance is achieved when the coil inductance is in the range of 40 nH to 400 nH.

**Calculation of Coil Magnetic Field**

Once the coil geometry is established and the coil current is simulated, the coil magnetic field can be calculated. This task is not trivial and usually requires numerical simulations. However, for the simple coil geometries, and if the coil shape is preserved on the time scale of the peak current, the coil magnetic field can be approximated with a simple formula. That is a good assumption for coils with inductance below 400 nH, because, as was already discussed, the time-to-peak current for such coils is below 2 µs, and the coil movement can be neglected. In the case of $N$-turn coil, the peak magnetic field at the coil center can be approximated by simple formula $B_0 = \mu_0 N I / d$, where $I$ is the coil current, $N$ is the number of the coil turns, and $d$ is the coil diameter. In this case, the magnetic field waveform, $B(t)$, is simply the scaled function of the current waveform, *I(t)*, which can be readily calculated with SCREAMER, or any other circuit code.

Here we aim to calculate the magnetic field generated inside a typical coil and compare the coil performance with the present, MIFEDS-2, pulser. The most direct approach to compare MIFEDS-2 and MIFEDS-3 pulsers is without the matched load, $\sqrt{L/C}$, assumption. In this case, the pulser total resistance is a sum of its internal value (130 mΩ for MIFEDS-2, and 140 mΩ for MIFEDS-3) and the resistance of the coil. Once again, we consider a simple, 2-turn coil with average coil diameter of 7.8 mm made of 24-AWG copper insulated wire. The inductance and resistance of this coil was calculated to be 60-nH and 140mΩ, respectively. The rest of the pulser parameters are as discussed in the introduction.

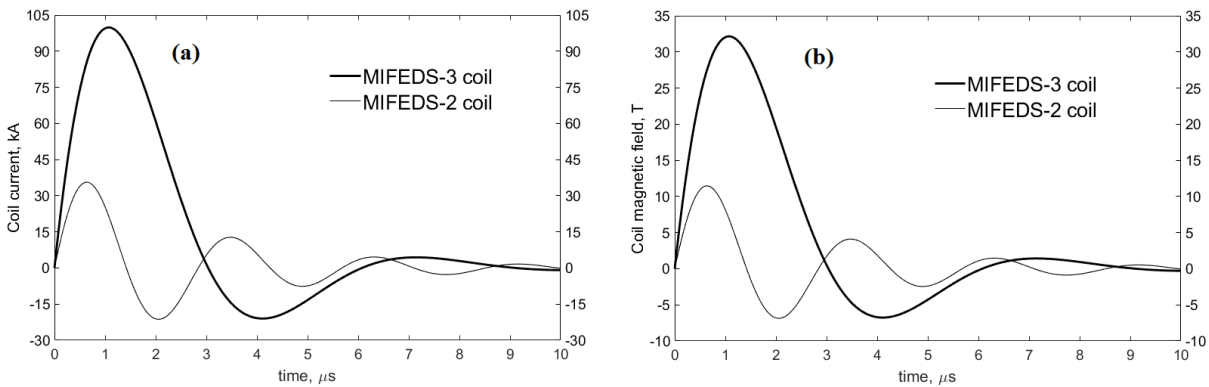

Figure 5. The currents and magnetic fields waveforms into the 60-nH, 14-mΩ coil. Fig. 5a corresponds to the MIFEDS coil current and Fig. 5b shows the peak magnetic field in the coil.

Both the current and magnetic field calculations into a 2-turn, 7.8-mm diameter coil are presented Figure *5*. As can be seen, for the new MIFEDS-3 pulser, the simulations predict maximum 100-kA peak current at about 1.1 µs. Not a surprise, this is factor of three larger than MIFEDS-2 current. It is interesting to observe, that the MIFEDS-3 current waveform is very close to the matched load case presented in Figure *2*. The slight difference in the peak and time-to-peak values account for the driver resistance mismatch as compared to the matched load. The coil magnetic fields (Figure *5*, b) are simply the scaled waveforms of the coil currents. The peak magnetic field generated inside the MIFEDS-3 coil reaches 32 T by the time of the current peak (1.1 µs), which is, as expected, the factor of three larger as compared to the present MIFEDS-2 pulser.

The full-width-half- maximum (FWHM) for MIFEDS-3 current and magnetic field waveforms is about twice as large as the MIFEDS-2 driver. In magnetized, high-energy-density experiments on OMEGA, the lasers are triggered at the current peak[12], a time interval where the current change is limited to ±5% (also called the flat-top window). The flat-top window is about 770 ns, relaxing the allowable jitter in triggering MIFEDS-3, compared to MIFEDS 2, which has a 380-ns window.

In experiments calling for a stronger magnetic field, it is possible to increase the total number of turns in the coil, keeping the coil inductance the same by reducing the coil diameter. It is also possible to increase the coil magnetic field keeping the coil diameter the same but increasing the total number of turns. Both cases are equally possible and result in the larger coil magnetic field. However, in the first case, the resulting magnetic field is concentrated over a small coil volume, and in the former case, the magnetic field is applied to the larger coil volume.

Here, we consider the generation of higher magnetic fields using four different coils, all having the same coil diameter, but with 1, 2, 3, and 4 total number of turns. To be consistent with our previous simulations, we assume all coils made of the same 24-AWG copper-insulated wire with 7.8-mm average coil diameter. The inductance and resistance of these coils are calculated utilizing the same approach used for the 2-turn coil. The coil inductances are 15 nH, 60 nH, 135 nH, and 240 nH with resistances of 12 mΩ, 14 mΩ, 16 mΩ, and 18 mΩ, for coils having 1, 2, 3, and 4 turns, respectively. Using SCREAMER, we simulate the current $I(t)$ waveform, and, later, using the formula for magnetic field $B_0 = \mu_0 NI/d$, calculate the resulting coil magnetic field, $B(t)$. The current waveform simulations for coils with different number of turns are presented in Figure 6a and the corresponding magnetic field waveforms are plotted in Figure 6b.

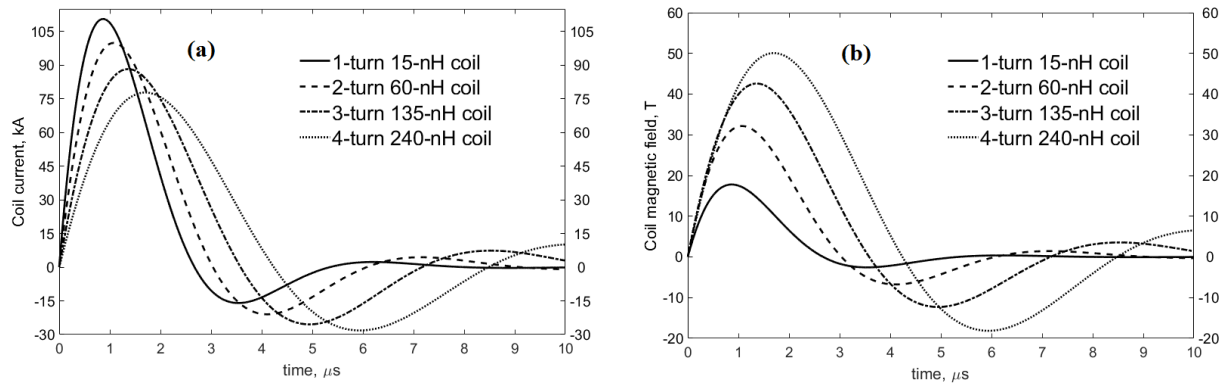

Figure 6. MIFEDS-3 simulations into 1-, 2-, 3- and 4-turn coils. All coils have same 7.8-mm average coil diameter. Right plot is the SCREAMER current waveform simulations, and the left plot is the corresponding coil magnetic field waveforms.

It is not obvious that an increase in the total number of turns, $N$, will result in a larger magnetic field. Indeed, as we increase the total number of turns, the coil inductance increases, and the coil peak current, $I(t)$, decreases. Because the magnetic field, $B(t)$, is proportional both to $I(t)$ and $N$, the resulting magnetic field is not obvious. As can be seen from Figure 6, left plot, the peak current gradually decreases as we increase the coil inductance. There is as much as 110-kA peak current with a 1-turn, 15-nH coil. The current decreases to about 80 kA for 4-turn, 240-nH coil. The peak magnetic fields monotonically increase as we increase the total number of turns. The peak magnetic fields are 18 T, 32 T, 42 T, and 50 T for the coils having 1, 2, 3, and 4 turns, correspondingly. It appears that the coil current, $I(t)$, is a weaker function than the total number of turns, $N$, and the resulting magnetic field become larger as we increase the coil number of turns.

It should be also noted, that as we increase the number of turns, the time-to-peak current increases as well. It appears that 1-4-turns are the optimal number of coil windings for the case with 7.8-mm average coil diameter. A larger number of turns will increase the time-to-peak current above 2 µs, which will result in motion of the coil wires during the pulse and a degradation of the peak magnetic field.

We finally present the comparison of the peak magnetic fields for different coils with the present MIFEDS-2 pulser. The peak magnetic fields for the new MIFEDS-3 pulser are from Figure 6, left plot. And the peak magnetic field for the present MIFEDS-2 pulsers are calculated for the same coils inductances (15 nH, 60 nH, 135 nH, and 240 nH). The result of this comparisons is presented in Figure 7. Not a surprise, the peak magnetic fields generated with the new pulser are the factor of three larger than peak magnetic fields generated inside the present MIFEDS-2 system. The increase in the peak magnetic fields is directly related to the capabilities of the upgraded pulser to supply the larger currents into some predefine magnetic coil.

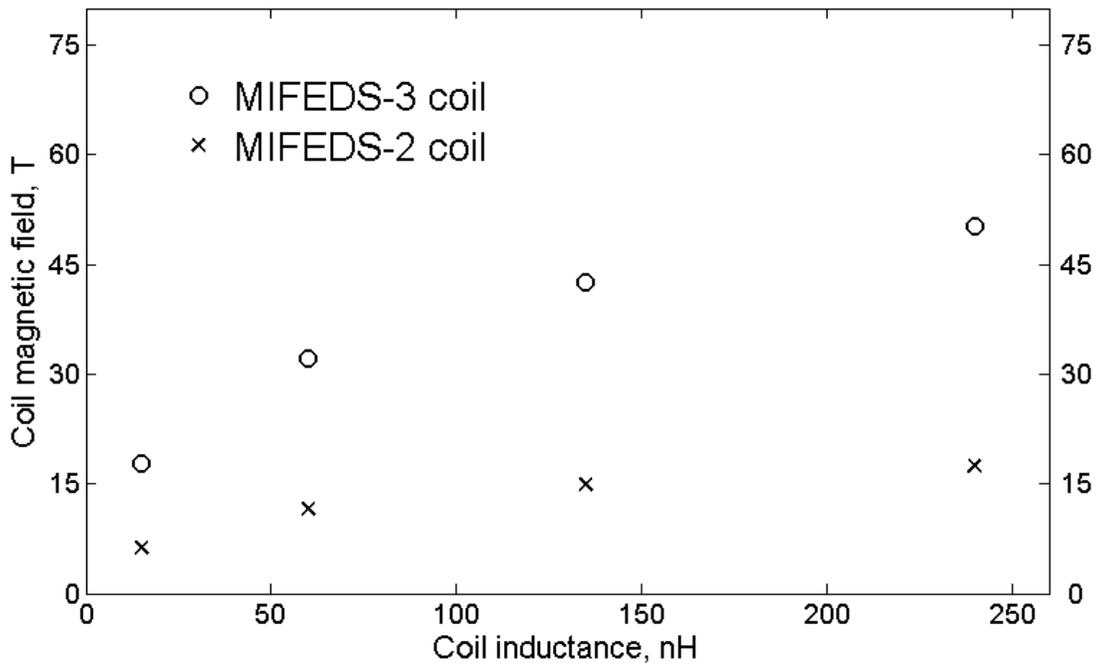

Figure 7. Peak magnetic fields are plotted as function of coil inductance. All coils have the same average coil diameter of 7.8 mm. The coil inductances are 15 nH, 60 nH. 135 nH, and 240 nH for 1-turn, 2-turns, 3-turns, and 4-turn coils, respectively.

**Conclusion**

We describe the electrical design of the new MIFEDS-3 pulser. By increasing the current capabilities by a factor of three, the proposed system can achieve stronger magnetic fields, while keeping the weight and volume of the system constant. The MIFEDS-3 design has a total stored electrical energy ten times larger than its predecessor and should reach 100-kA peak current in less than 1 μs, into 60-nH, 0.14-mΩ coil. Assuming a 2-turn coil geometry with an average coil diameter of 7.8 mm this peak current gives a peak magnetic field of 32 T at the center of the coil. Even larger magnetic fields can be achieved using coils with more turns. We believe that MIFEDS-3 gives the best magnetic-field performance possible on OMEGA given the volume and weight limitations imposed on the pulser by the TIM.